\documentclass[apj]{emulateapj}

\usepackage{natbib}
\usepackage{color}
\usepackage{mediabb}
\usepackage{amsmath,amssymb}

\newcommand{\myemail}{hideki.umehata@ouj.ac.jp}

\slugcomment{draft November 26 2016}

\shorttitle{[\textsc{Cii}] from LAB1}
\shortauthors{Umehata et al.}

\begin{document}


\title{
ALMA reveals Strong [CII]  emission in a galaxy embedded in a giant Ly$\alpha$ blob at $z=3.1$
}


\author{
Hideki Umehata\altaffilmark{1,2}, Yuichi Matsuda\altaffilmark{3,4}, Yoichi Tamura\altaffilmark{2}, Kotaro Kohno\altaffilmark{2,5}, Ian Smail\altaffilmark{6}, R.J. Ivison\altaffilmark{7,8}, Charles~C.~Steidel\altaffilmark{9}, Scott C. Chapman\altaffilmark{10}, James E. Geach\altaffilmark{11}, Matthew Hayes\altaffilmark{12}, Tohru Nagao\altaffilmark{13}, Yiping Ao\altaffilmark{3}, Ryohei Kawabe\altaffilmark{3,4,15}, Min S. Yun\altaffilmark{14}, Bunyo Hatsukade\altaffilmark{3}, Mariko Kubo\altaffilmark{3}, Yuta Kato\altaffilmark{3,15}, Tomoki Saito\altaffilmark{16}, Soh Ikarashi\altaffilmark{17}, Kouichiro~Nakanishi\altaffilmark{3,4}, Minju Lee\altaffilmark{3,15}, Takuma Izumi\altaffilmark{2}, Masao~Mori\altaffilmark{18}, Masami Ouchi\altaffilmark{19}
}
\affil{
$^1$ The Open University of Japan, 2-11 Wakaba, Mihama-ku, Chiba 261-8586, Japan; \textcolor{blue}{{\myemail}}
\\
$^2$ Institute of Astronomy, School of Science, The University of Tokyo, 2-21-1 Osawa, Mitaka, Tokyo 181-0015, Japan
\\
$^3$ National Astronomical Observatory of Japan, 2-21-1 Osawa, Mitaka, Tokyo 181-8588, Japan
\\
$^4$ Department of Astronomy, School of Science, SOKENDAI (The Graduate University for Advanced Studies), Osawa, Mitaka,
Tokyo 181-8588, Japan
\\
$^5$ Research Center for the Early Universe, The University of Tokyo, 7-3-1 Hongo, Bunkyo, Tokyo 113-0033
\\
$^6$ Centre for Extragalactic Astronomy, Department of Physics, Durham University, South Road, Durham, DH1 3LE, UK
\\
$^7$ European Southern Observatory, Karl-Schwarzschild-Str. 2, D-85748 Garching, Germany
\\
$^8$ Institute for Astronomy, University of Edinburgh, Royal Observatory, Blackford Hill, Edinburgh EH9 3HJ, UK
\\
$^{9}$ California Institute of Technology, MS 249-17, Pasadena, CA 91125, USA
\\
$^{10}$ Department of Physics and Atmospheric Science, Dalhousie University, Halifax, NS B3H
4R2, Canada
\\
$^{11}$ Centre for Astrophysics Research, Science \& Technology Research Institute, University of
Hertfordshire, Hatfield AL10 9AB, UK
\\
$^{12}$ Department of Astronomy, Oskar Klein Centre, Stockholm University, AlbaNova University
Centre, SE-106 91 Stockholm, Sweden
\\
$^{13}$ Research Center for Space and Cosmic Evolution, Ehime University, 2-5 Bunkyo-cho,
Matsuyama, Ehime 790-8577
\\
$^{14}$ Department of Astronomy, University of Massachusetts, Amherst, Massachusetts 01003
\\
$^{15}$ Department of Astronomy, Graduate school of Science, The University of Tokyo, 7-3-1
Hongo, Bunkyo-ku, Tokyo 133-0033
\\
$^{16}$ Nishi-Harima Astronomical Observatory, Centre for Astronomy,
University of Hyogo, 407-2 Nichigaichi, Sayo-cho, Sayo,
Hyogo 679-5313, Japan
\\
$^{17}$ Kapteyn Astronomical Institute, University of Groningen, P.O. Box 800, 9700AV Groningen, The Netherlands
\\
$^{18}$ Center for Computational Physics, University of Tsukuba, 1-1-1 Tennodai, Tsukuba, Ibaraki
305-8577
\\
$^{19}$ Institute for Cosmic Ray Research, University of Tokyo, 5-1-5 Kashiwa-no-Ha, Kashiwa City,
Chiba 277-8582
}

%




\begin{abstract}
We report the result from observations conducted with the Atacama Large Millimeter/submillimeter Array (ALMA) to detect [\textsc{Cii}] 158~$\mu$m fine structure line emission from galaxies embedded in one of the most spectacular Ly$\alpha$ blobs (LABs) at $z=3.1$, SSA22-LAB1.
Of three dusty star-forming galaxies previously discovered by ALMA 860~$\mu$m dust continuum survey toward SSA22-LAB1, we detected the [\textsc{Cii}] line from one, LAB1-ALMA3 at $z$=3.0993$\pm$0.0004.
No line emission was detected, associated with the other ALMA continuum sources or from three 
rest-frame UV/optical selected $z_{\rm spec}\simeq3.1$ galaxies within the field of view.
For LAB1-ALMA3, we find relatively bright [\textsc{Cii}] emission compared to the infrared luminosity ($L_\textsc{[Cii]}$/$L_{\rm IR}\approx0.01$) and 
an extremely high [\textsc{Cii}] 158~$\mu$m and [\textsc{Nii}] 205~$\mu$m emission line ratio
($L_\textsc{[Cii]}$/$L_\textsc{[Nii]}>$55).
The relatively strong [\textsc{Cii}] emission may be caused by abundant photodissociation regions and sub-solar metallicity, or by
 shock heating.
The origin of the unusually strong [\textsc{Cii}] emission could be causally related to the location within the giant LAB, although the relationship between extended Ly$\alpha$ emission and ISM conditions of associated galaxies is yet to be understand.
 \end{abstract}





\section{Introduction}

\begin{figure*}
\epsscale{1.15}
\plotone{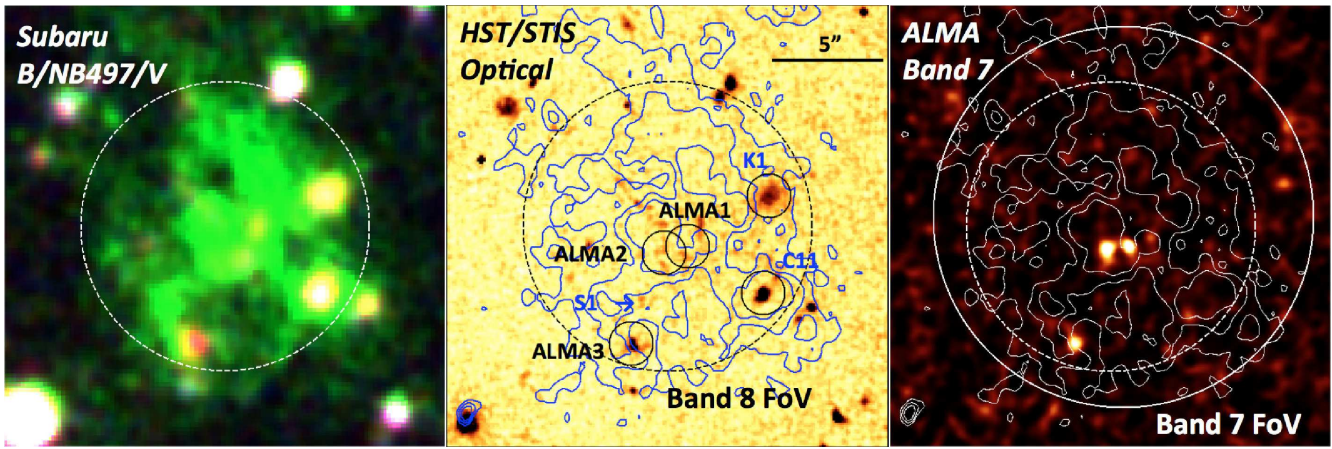}
\caption{
Images of SSA22-LAB1. Each field is $20^{\prime\prime}\times20^{\prime\prime}$ in size. The field of view of the ALMA band~8 observation is shown in each figure.
({\it left}) A pseudo color image with Subaru/Suprime-Cam $B$-, $NB497$-, and $V$-band where the strong Ly$\alpha$ emission falls in the green channel (\citealt{2004AJ....128..569M}). 
({\it middle}) {\it HST} STIS optical image as a finding chart. Contours show Ly$\alpha$ emission at levels of 4, 8, and 12~$\times10^{-18}$~erg~s$^{-1}$~cm$^{-2}$~arcsec$^{-2}$ (\citealt{2004AJ....128..569M}). We show the positions of three ALMA sources (ALMA1, ALMA2, and ALMA3; \citealt{2016ApJ...832...37G}; Y. Matsuda et al. in preparation) and other $z_{\rm spec}\approx3.1$ galaxies: one LBG (C11; \citealt{2003ApJ...592..728S}), and one {\it K}-selected galaxy (K1; \citealt{2015ApJ...799...38K}).
One faint [\textsc{Oiii}] emitter at $z=3.0968$ (S1; \citealt{2016ApJ...832...37G}) is also shown.
({\it right}) The non-primary-beam-corrected ALMA image at 860~$\mu$m (Y. Matsuda et al. in preparation).
}
\label{f1}
\end{figure*}

Investigating the physical and chemical properties of the interstellar medium (ISM) of dusty star-forming galaxies and/or high-redshift galaxies has been difficult, as typical UV/optical nebular lines are not useful due to heavy dust extinction and/or the lines are not accessible with conventional ground-based instruments.
Recently, the Atacama Large Millimeter/submillimeter Array (ALMA) has opened a new window, allowing us to exploit fine structure lines at rest-frame far-infrared (FIR) wavelengths to diagnose the ISM properties for these galaxy populations (e.g., \citealt{2012A&A...542L..34N}; \citealt{2014ApJ...782L..17D}; \citealt{2016Sci...352.1559I}).
The [\textsc{Cii}]~158~$\mu$m ($^2P_{3/2}\rightarrow^2P_{1/2}$) is known to be the dominant coolant of the ISM and one of the brightest lines from star-forming galaxies in the FIR (e.g., \citealt{1996A&A...308..723I}).
While the [\textsc{Cii}] emission arises primarily from dense photodissociation regions (PDRs),  it is also observed in various regions/environments, including ionized regions, cool, diffuse interstellar gas, and shocked gas (e.g., \citealt{1991ApJ...373..423S}; \citealt{1993ApJ...407..579M}; \citealt{2011A&A...526A.149N}; \citealt{2013ApJ...777...66A}). 

In order to characterize the [\textsc{Cii}] emission and investigate the nature of the ISM in star-forming galaxies at high redshift,
Ly$\alpha$ blobs (LABs) are a useful laboratory.
LABs are extended gaseous nebulae, preferentially found in regions of galaxy overdensities in the distant universe (e.g., \citealt{2000ApJ...532..170S}; \citealt{2004AJ....128..569M}; \citealt{2009ApJ...693.1579Y}).
A large number of LABs are associated with star-forming galaxies such as submillimeter galaxies (SMGs; e.g., \citealt{2005MNRAS.363.1398G}, \citeyear{2014ApJ...793...22G}; \citealt{2015ApJ...815L...8U}, \citeyear{2016arXiv161109857U}), distant red galaxies (DRGs; e.g., \citealt{2011ApJ...740L..31E}; \citealt{2012ApJ...750..116U}; \citealt{2013ApJ...778..170K}), and Lyman break galaxies (LBGs; e.g., \citealt{2004AJ....128..569M}).
Thus LABs are likely to be the sites of ongoing massive galaxy formation and assembly, and the extended gaseous structures around them are believed to be observational signs of large-scale gas flows (inflow/outflow) and their interactions as well as photoionization (e.g., \citealt{2000ApJ...532L..13T}; \citealt{2006Natur.440..644M}; \citealt{2009MNRAS.400.1109D}).
%
SSA22-LAB1 (hereafter LAB1, \citealt{2000ApJ...532..170S}) is a giant LAB discovered in the $z=3.1$ SSA22 proto-cluster region and one of the most well-studied LABs (e.g., \citealt{2004ApJ...606...85C}; \citealt{2014ApJ...793...22G}; \citealt{2011Natur.476..304H}; \citealt{2015ApJ...799...38K}).
The unique environment
makes LAB1 a useful laboratory for investigating the [\textsc{Cii}] emission from growing galaxies in the early universe.
Throughout the paper, we adopt a cosmology with 
$\Omega_{\rm m}=0.3, \Omega_\Lambda=0.7$, and H$_0$=70 km s$^{-1}$ Mpc$^{-1}$.

\section{Observations and data reduction}

We observed LAB1 with ALMA in band~8 as a part of an ALMA cycle-2 program (ID: 2013.1.00159.S; PI: Umehata), targeting the \textsc{[Cii]} 158~$\mu$m transition ($\nu_{\rm rest}=1900.537$~GHz, redshifted to 463.55~GHz or 647~$\mu$m, at $z=3.100$).
As shown in Fig.~\ref{f1}, the field of view (FoV) at $\sim$ 464 GHz is large enough to cover the majority of the Ly$\alpha$ emitting region ($d\sim13.5^{\prime\prime}$ or $\sim 100$ kpc at $z=3.1$ ) and contains three $860~\mu$m continuum ALMA sources: LAB1-ALMA1, LAB1-ALMA2, and LAB1-ALMA3 (hereafter ALMA1, ALMA2, and ALMA3, respectively; \citealt{2016ApJ...832...37G})\footnote{ALMA1, ALMA2, and ALMA3 correspond to SSA22-LAB01 ALMA b, SSA22-LAB01 ALMA a, and SSA22-LAB01 ALMA c in \citet{2016ApJ...832...37G}, respectively.}.
ALMA3 is spatially coincident with a DRG at $z_{\rm spec}=3.1$ (\citealt{2015ApJ...799...38K}).
While ALMA1 and ALMA2 do not have spectroscopic redshifts, their photometric redshifts and 
the low probability of chance association of ALMA sources suggest a physical association between the two ALMA sources and the giant Ly$\alpha$ nebula (\citealt{2012ApJ...750..116U}; Y. Matsuda et al. in preparation).
Three other galaxies at $z_{\rm spec}\simeq3.1$ (a LBG, a K-band~selected galaxy, and a [\textsc{Oiii}] emitter) are also located within the band~8 FoV (Fig.~\ref{f1}).

\begin{figure*}
\epsscale{1.16}
\plotone{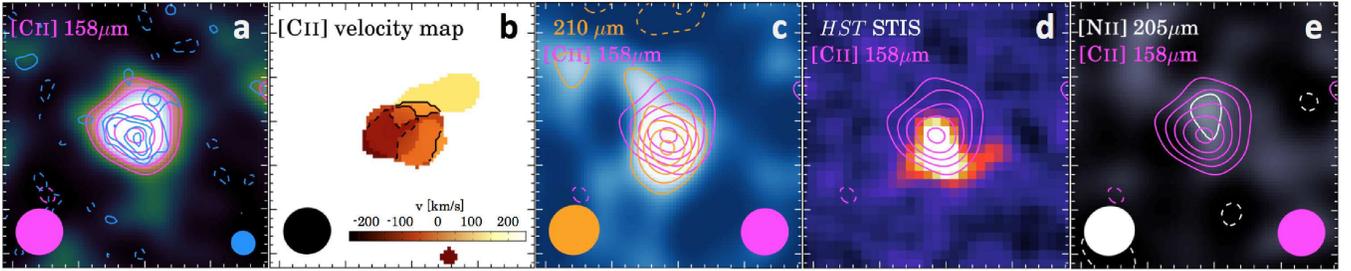}
\caption{
Images of LAB1-ALMA3. The size of each map is $3^{\prime\prime}\times3^{\prime\prime}$.
(a) The velocity-integrated map of the [\textsc{Cii}] emission. The background map is the ``tapered'' map (0$^{\prime\prime}$.53 FWHM, magenta contours), while we also show the ``full'' map (0$^{\prime\prime}$.27 FWHM, blue contours) for comparison. Contours start at $\pm2\sigma$, with steps of 1$\sigma$ for both.
(b) The velocity map of the  [\textsc{Cii}] emission, blanked at 2.5$\sigma$.
Velocities are relative to the [\textsc{Oiii}] peak (see also Fig. \ref{spectra}) and velocity contours are shown in steps of 80~km s$^{-1}$.
(c) The ``tapered'' band~7 continuum map (0$^{\prime\prime}$.55 FWHM), which presents rest-frame 210~$\mu$m continuum emission.
Contours are plotted from $\pm2\sigma$ in steps of 1$\sigma$.
For comparison, we also show contours of the tapered [\textsc{Cii}] map presented in panel a.
(d) The $HST$ STIS optical image, compared to the [\textsc{Cii}] emission.
(e) The ``tapered'' [\textsc{Nii}] map. 
Contours are $\pm2\sigma$.
[\textsc{Cii}] emission is same as other panels.
}
\label{cii_map}
\end{figure*}

\begin{figure}
\epsscale{1.15}
\plotone{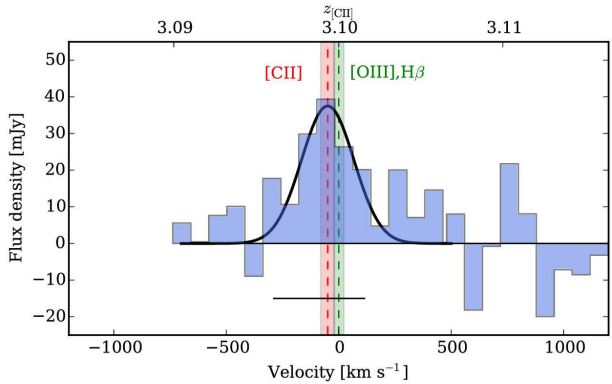}
\caption{
[\textsc{Cii}] spectrum of LAB1-ALMA3, integrated over a region of $d=1^{\prime\prime}$ in the tapered cube after correcting for the primary-beam response.
We also show the redshifts and errors determined from [\textsc{Cii}] (red lines) and [\textsc{Oiii}]/H$\beta$ (green lines) detections.
Velocities are relative to the [\textsc{Oiii}]/H$\beta$ redshift ($z=3.1000\pm0.0003$; \citealt{2015ApJ...799...38K}).
[\textsc{Cii}] emission from ALMA3 is detected at consistent redshift ($z=3.0993\pm0.0004$) with FWHM of $270\pm30$ km~s$^{-1}$.
The velocity range used to create the images in Fig.~\ref{cii_map} is indicated below the spectrum.
}
\label{spectra}
\end{figure}

Observations were carried out on 16 June 2015 using a spectral scan mode with the FDM correlator mode to cover the redshift range of the proto-cluster, $z=3.06-3.12$ (\citealt{2005ApJ...634L.125M}).
Among four planned spectral windows, only two were actually executed.
The incomplete observation resulted in frequency coverage of 461.03--462.78 GHz and 462.91--464.66 GHz ($z_\textsc{[Cii]}$=3.090--3.105, 3.107--3.122) after flagging the edge channels.
The array configuration was C34-5 and the baseline lengths were 21--784~m.
The on-source time was 4.5~minutes.
Ceres was observed for amplitude calibration, and 
the quasar J2148+0657 was utilized for bandpass and phase calibration.
The data were processed with the Common Astronomy Software Application ({\sc casa}) ver.~4.4.0 (\citealt{2007ASPC..376..127M}).
The cube was first created with the natural weighting using the {\sc casa} task, {\sc clean}.
The resultant cube (hereafter ``full'' cube) has a typical synthesized beam FWHM of $0^{\prime\prime}.27\times0^{\prime\prime}.26$ (P.A. 46 deg).
We also created a ``tapered'' cube adopting the taper parameter, outertaper = 0.5 arcsec, which has a typical synthesized beam, $0^{\prime\prime}.53\times0^{\prime\prime}.52$ (P.A. $-70$ deg).
The typical rms level is $\approx$ 3.5 mJy beam$^{-1}$ at the phase center per 80 km s$^{-1}$ channel in the tapered cube. 
To search for band~8 continuum sources, we created a ``tapered'' continuum map at 463~GHz, using the line-free channels.
The ``dirty'' map has a rms level of 0.8~mJy beam$^{-1}$ at the phase center and none of the sources is found above 5$\sigma$. 

LAB1 has also been observed by ALMA in band~7.
One program (ID. 2013.1.00704.S; PI. Matsuda) covered the redshifted \textsc{[Nii]} 205~$\mu$m transition line ($\nu_{\rm rest}=1461.131$~GHz, redshifted to 356.37~GHz, at $z=3.100$) (Y. Matsuda et al. in preparation). 
The typical noise rms at $0^{\prime\prime}.55$ resolution, which is equivalent to the ``tapered'' cube in band~8, is $\approx$0.4 mJy beam$^{-1}$ at the phase center, per 80 km s$^{-1}$ channel. 

\section{Results}

\subsection{\textsc{[Cii]} 158~$\mu$m in LAB1-ALMA3}

We detected \textsc{[Cii]} emission from one of the three dusty star-forming galaxies, ALMA3 (Fig.~\ref{cii_map} and Fig.~\ref{spectra}).
Fig.~\ref{spectra} shows the \textsc{[Cii]} spectrum.
A gaussian fit to the line has $z=3.0993\pm0.0004$ with FWHM $275\pm30$ km~s$^{-1}$.
\citet{2015ApJ...799...38K} reported a redshift of $z=3.1000\pm0.0003$ on the basis of H$\beta$ and \textsc{[Oiii]} $\lambda$5007 lines, and hence our measurement is  consistent (the velocity offset is within $\sim50$~km~s$^{-1}$ and the two measurements are consistent within errors).
Fig.~\ref{cii_map} shows the velocity-integrated \textsc{[Cii]} intensity and velocity maps, compared to the rest-frame 210~$\mu$m continuum (Y. Matsuda et al. in preparation; \citealt{2016ApJ...832...37G}),  $HST$ STIS optical image\footnote{The image has a pivot wavelength of 5733~\AA.} (\citealt{2003ApJ...599...92C}), and \textsc{[Nii]} image\footnote{We created the \textsc{[Nii]} image, integrated the cube over the same velocity range of the \textsc{[Cii]} map.}.
The \textsc{[Cii]} emission is spatially resolved as shown in Fig.~\ref{cii_map}a, while
the \textsc{[Cii]} emission has a modest signal to noise ratio and the various clumps seen are not significant.
The \textsc{[Cii]} velocity map (Fig. 2b) also shows complexity, which is not likely to be produced by a simple rotating disk.
The position of \textsc{[Cii]} emission is generally consistent with those of dust continuum and stellar emission\footnote{There might be a small offset, $\sim0^{\prime\prime}.2$, though the current data is insufficient to determine whether it is real.}.

\begin{center}
\begin{deluxetable*}{ccccccccccc}
\tabletypesize{\scriptsize}
\tablecaption{[\textsc{Cii}] Line Parameters of galaxies in SSA22-LAB1}
\tablewidth{0pt}
\tablehead{
\colhead{Galaxy} &\colhead{RA} &\colhead{Dec} & \colhead{$z$} & \colhead{Type} & \colhead{Ref} & $I_\textsc{Cii}$ &  $L_\textsc{Cii}$ & $L_{\rm IR}$ \\
                             & (J2000)        & (J2000)              &            &                           &                              &   (Jy km$^{-1}$)     & (10$^{9} L_\sun$) & (10$^{11} L_\sun$)  \\                 
}
\startdata
LAB1-ALMA3 & 22:17:26.11 &   +00:12:32.4 & $3.0993\pm0.0004$  & [\textsc{Cii}] 158~$\mu$m                     & 1   & 16.8$\pm$2.1 & 5.7$\pm0.7$ & 5.8 \\
                       &  22:17:26.1  &  +00:12:32.3  & $3.1000\pm0.0003$  & [\textsc{Oiii}] $\lambda5007$, H$\beta$ & 2 &       ---        & --- & \\ 
LAB1-ALMA1 & 22:17:25.94  &  +00:12:36.6 & (3.1?) & photo-$z$ & --- & ($<2.3$) & ($<0.8$) & 3.5 \\
LAB1-ALMA2 & 22:17:26.01 &   +00:12:36.4 & (3.1?) &  photo-$z$ & --- & ($<2.3$) & ($<0.8$) & 4.0 \\
C11 (LBG)     & 22:17:25.7  &  +00:12:34.7 & $3.0999\pm0.0004$ & [\textsc{Oiii}] $\lambda5007$ & 3 &  $<3.0$ & $<1.0$ & --- \\
K1 (K-band~galaxy) & 22:17:25.7 &   +00:12:38.7 & $3.1007\pm0.0002$ & [\textsc{Oiii}] $\lambda5007$ & 2 & $<2.6$ & $<0.9$& ---\\
S1 ([\textsc{Oiii}] emitter) & 22:17:26.08  &  +00:12:34.2 &   3.0968 & [\textsc{Oiii}] $\lambda5007$ & 4 &$<2.2$ & $<0.7$ & ---
\enddata
\tablecomments{
[\textsc{Cii}] Line properties of three ALMA sources and three UV/optical selected galaxies.
Since ALMA1 and ALMA2 don't have $z_{\rm spec}$, we estimated rough upper limits using the cube for ALMA3, assuming same redshifts and velocity widths.
For C11, K1, and S1, we integrated the cube at the position in literatures over $300$~km s$^{-1}$ velocity range, and obtain 3$\sigma$ upper limits.
References are: 1. This work, 2. \citealt{2015ApJ...799...38K}, 3. \citealt{2013ApJ...767...48M}, and 4. \citealt{2016ApJ...832...37G}.
}
\end{deluxetable*}
\end{center}

To describe the properties of \textsc{[Cii]} emission from the whole galaxy, we use the tapered map.
A two-dimensional elliptical Gaussian fit yields a deconvolved FWHM of ($0^{\prime\prime}.62\pm0^{\prime\prime}.11)\times(0^{\prime\prime}.55\pm0^{\prime\prime}.10$), which corresponds to $4.8\times4.3$~kpc$^2$.
For comparison, we similarly measured the size of the dusty starburst core using the band~7 continuum image at 0$^{\prime\prime}$.35 resolution.
The yielded size is ($0^{\prime\prime}.53\pm0^{\prime\prime}.14)\times(0^{\prime\prime}.40\pm0^{\prime\prime}.12$) ($4.1\times3.1$~kpc$^2$).
The measured integrated line flux is $I_\textsc{[Cii]}=16.8\pm2.1$~Jy km s$^{-1}$ and hence the line luminosity is $L_\textsc{[Cii]}=(5.7\pm0.7) \times10^{9} L_\sun$ (Table 1). 
The infrared (IR; 8-1000~$\mu$m) luminosity of ALMA3 is derived using an average SMG template from the ALESS survey (\citealt{2014MNRAS.438.1267S}) scaled to the 860~$\mu$m flux density, $S_{\rm 860 \mu m}=0.73\pm0.05$~mJy (\citealt{2016ApJ...832...37G}); $L_{\rm IR}\approx5.8\pm0.4\times10^{11} L_\odot$, so that $L_\textsc{[Cii]}/L_{\rm IR}\approx0.010\pm0.001$
(We note that the IR luminosity may have larger uncertainty.
\citet{2016ApJ...832...37G} estimated it in the range $L_{\rm IR}\approx(0.2-1.5)\times10^{12} L_\odot$ using varying templates.).
We also derived the dynamical mass of ALMA3, $M_{\rm dyn, vir}\sim 1.0\times10^{11} M_\odot$, using an isotropic virial estimator (e.g., \citealt{2010ApJ...724..233E}) on the basis of the line width and \textsc{[Cii]} size (major axis measured from the FWHM).

We also searched for \textsc{[Nii]} 205~$\mu$m emission from ALMA3, which resulted in non-detection (Fig.~\ref{cii_map}e).
Utilizing the \textsc{[Nii]} map at $0^{\prime\prime}.55$ resolution, we obtained a 3$\sigma$ (point-source) upper limit on its line intensity, $I_\textsc{[Nii]}<0.35$~Jy km s$^{-1}$ and 
thus $L_\textsc{[Nii]}<9.4\times10^{7} L_\sun$, and $L_\textsc{[Cii]}/L_\textsc{[Nii]}>61$.
The \textsc{[Nii]} upper limit can slightly be relaxed when the \textsc{[Nii]} 205~$\mu$m emission has larger extent compared to the size of the synthesized beam.
If we use the the other tapered \textsc{[Nii]} map at $0^{\prime\prime}.64$ resolution, which is comparable to the measured \textsc{[Cii]} size of ALMA3, we will have $I_\textsc{[Nii]}<0.39$~Jy km s$^{-1}$, $L_\textsc{[Nii]}<1.0\times10^{8} L_\sun$, and $L_\textsc{[Cii]}/L_\textsc{[Nii]}>55$, respectively.
In the following discussion, we adopt the latter conservatively.

\subsection{No \textsc{[Cii]} emission from the remaining LAB1 members}

Except for ALMA3, no emission line is found in the band~8 cube.
For ALMA1 and ALMA2, we just calculate a tentative upper limit of \textsc{[Cii]} emission, assuming that 
the lines fall within our frequency coverage and the line widths are same as that of ALMA3.
The IR luminosities of ALMA1 and ALMA2 are comparable to that of ALMA3 ($L_{\rm IR}\approx3.5\times10^{11}L_\odot$ and $L_{\rm IR}\approx4.0\times10^{11}L_\odot$, respectively)\footnote{
\citet{2016ApJ...832...37G} reported the sum of 860~$\mu$m flux density, $S_{\rm 860 \mu m}=0.95\pm0.04$~mJy.
We apportioned it between ALMA1 and ALMA2 according to their peak flux density at $0^{\prime\prime}.35$ resolution (Y. Matsuda et al. in preparation) and calculated IR luminosity in the same way for ALMA3.}.
Utilizing the intensity map for ALMA3, we obtained a 3$\sigma$ upper limit on their individual line intensity, $I_\textsc{[Cii]}<2.3$~Jy km s$^{-1}$, and line luminosity, $ L_\textsc{[Cii]}<0.8 \times10^{9} L_\sun$.
Although this is just a crude estimate and $z_{\rm spec}$ information is essential for further discussion, 
our result suggests that the $L_\textsc{[Cii]}/L_{\rm IR}$ of ALMA1 and ALMA2 may be different from that of ALMA3.
We also evaluated 3$\sigma$ upper limits for the three rest-frame UV/optical galaxies with \textsc{[Oiii]} line detections, by integrating the cube over 300~km~s$^{-1}$ at the source position (Table 1).

\section{Discussion and Summary}

\begin{figure*}
\epsscale{1.15}
\plotone{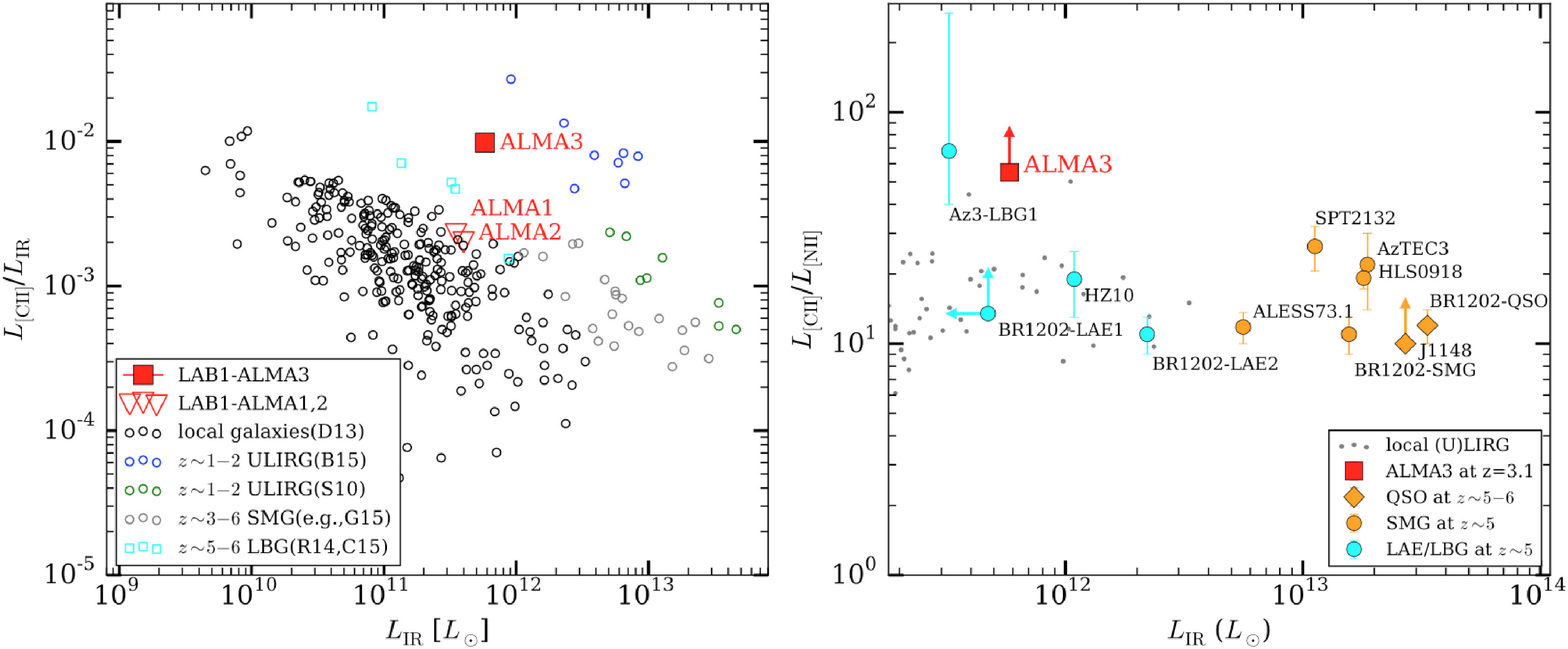}
\caption{
({\it left}) \textsc{[Cii]}-IR luminosity ratio ($L_\textsc{[Cii]}/L_{\rm IR}$) as a function of IR luminosity ($L_{\rm IR}$). 
We show the measured ratio of LAB1-ALMA3 and the ``upper limit'' of LAB1-ALMA1 and ALMA2, 
assuming their redshifts lie within our \textsc{[Cii]} coverage (see text).
We also mark local IR-luminous galaxies (\citealt{2013ApJ...774...68D}), the SMGs at $z=3\sim6$ (\citealt{2014ApJ...796...84R} (R14); \citealt{2014ApJ...782L..17D}; \citealt{2014A&A...565A..59D}; \citealt{2014ApJ...783...59R}; \citealt{2015MNRAS.449.2883G} (G15)), LBGs at $z=5-6$ (R14; \citealt{2015Natur.522..455C} (C15)), and $z=1\sim2$ star-forming galaxies (including SMGs; \citealt{2010ApJ...724..957S} (S10); \citealt{2015ApJ...799...13B} (B15)).
ALMA3 shows high \textsc{[Cii]}-IR luminosity ratio, compared to other IR luminous galaxies with similar luminosity. Here we convert the IR luminosities in the literature, multiplying by the following factors: $L_{\rm 8-1000\mu m}/L_{\rm 42.5-122.5\mu m}=1.7$, $L_{\rm 8-1000\mu m}/L_{\rm 42.5-500\mu m}=1.3$.
({\it right}) \textsc{[Cii]} 158~$\mu$m-\textsc{[Nii]} 205~$\mu$m line luminosity ratio ($L_\textsc{[Cii]}/L_\textsc{[Nii]}$) as a function of IR luminosity ($L_{\rm IR}$). 
The ratio of  LAB1-ALMA3 is shown, compared with those of various galaxies at $z\sim5$ (\citealt{2014ApJ...783...59R}; \citealt{2014ApJ...782L..17D}; \citealt{2016A&A...586L...7B}; \citealt{2016arXiv160702520P}, and references therein) and local (U)LIRGs (\citealt{2016ApJ...819...69Z}; \citealt{2013ApJ...774...68D}).
ALMA3 shows one of the highest values seen to date, which indicates an enhanced \textsc{[Cii]} emission.}
\label{ratio}
\vspace{0.79cm}
\end{figure*}

One striking characteristic of ALMA3 is the high \textsc{[Cii]}--IR ratio seen in Fig.~\ref{ratio}.
While this ratio is known to decrease as IR luminosity increases (``\textsc{[Cii]} deficit'') for local and high-redshift IR luminous galaxies (e.g., \citealt{2013ApJ...774...68D}), ALMA3 shows approximately an order of magnitude higher ratio (Fig.~\ref{ratio}) at the same IR luminosity range
(This trend is independent of the uncertainties on $L_{\rm IR}$ described in \S 3.1. 
While the $L_\textsc{[Cii]}/L_{\rm IR}$ ratio may be $\sim\times$3 lower, the increased corresponding $L_{\rm IR}$ keeps the trend.).
The result implies different conditions responsible for \textsc{[Cii]} emission between ALMA3 and the majority of previously known IR luminous galaxies.
It has also been reported that some $z\sim1-2$ ULIRGs show $L_\textsc{[Cii]}/L_{\rm IR}$ ratios comparable to ALMA3, although they have slightly higher $L_{\rm IR}$ than ALMA3 (\citealt{2015ApJ...799...13B}).
One possible explanation for elevated \textsc{[Cii]}--IR ratios is that the galaxies host widely spread star formation, and the UV radiation field is therefore diluted, which make the \textsc{[Cii]} line a more efficient coolant (see e.g., \citealt{2015A&A...574A..14C}; \citealt{2015ApJ...799...13B}, and references therein).
The size of the dust continuum core in ALMA3 is 4.1~kpc, which
is larger than a typical continuum size of bright SMGs at similar redshifts (2.4~kpc; \citealt{2015ApJ...799...81S}; see also \citealt{2015ApJ...810..133I}; \citealt{2016arXiv161109857U}). 
This supports that a relatively extended star-forming region in ALMA3 contributes the high \textsc{[Cii]}/IR ratio for ALMA3.
Gas accretion from the cosmic web is expected to accumulate a large amount of molecular gas necessary to fuel such widespread star formation (\citealt{2015ApJ...799...13B}).

We have another clue from the \textsc{[Cii]} 158~$\mu$m--\textsc{[Nii]} 205~$\mu$m line luminosity ratio, $L_\textsc{[Cii]}/L_\textsc{[Nii]}$.
ALMA3 shows one of the largest ratios ever reported (Fig.~\ref{ratio}).
The $L_\textsc{[Cii]}/L_\textsc{[Nii]}$ ratio has been utilized to diagnose the ISM conditions.
In particular, it is used to evaluate global trend on the fraction of \textsc{[Cii]} emission associated with ionized regions (i.e. \textsc{Hii} regions; e.g., \citealt{2006ApJ...652L.125O}; \citealt{2014ApJ...782L..17D};  \citealt{2016arXiv160702520P}), mainly because Nitrogen's ionization potential (14.5~eV) is higher than that of Hydrogen (13.6~eV) so that \textsc{[Nii]} arises only from ionized regions.
\citet{2016arXiv160702520P} reported the expected a line ratio $L_\textsc{[Cii]}/L_\textsc{[Nii]}\approx3.5$, for \textsc{Hii} regions with electron density of $\sim10-1000$~cm$^{-3}$.
If we adopt this estimate, it is expected that the contribution of ionized gas is only about $\sim$6\% and the vast majority of \textsc{[Cii]} emission arises from the surface of dense PDRs and/or other regions/environments.
The $L_\textsc{[Cii]}/L_\textsc{[Nii]}$ ratio is also sensitive to estimate gas metallicity (e.g., \citealt{2012A&A...542L..34N}; \citealt{2016A&A...586L...7B}; \citealt{2016arXiv160702520P}).
\citet{2012A&A...542L..34N} suggests that the line ratio increases as metallicity decreases, considering both PDRs and \textsc{Hii} regions in their model.
The measured ratio, $L_\textsc{[Cii]}/L_\textsc{[Nii]}>55$, favors sub-solar metallicity for the variety of densities and ionization parameters in their model. 
Gas accretion from the outside of ALMA3 may explain this relatively low metallicity.
It is suggested that Nitrogen may dominantly be in its doubly ionized state in high ionization conditions with lower dust shielding (e.g., \citealt{2016arXiv160702520P}).
This effect is unlikely to be significant in ALMA3 because it is detected in dust continuum.

Although it is not straightforward to identify the origin of \textsc{[Cii]} emission more, together with these clues, 
the properties and location of ALMA3 may support the importance of shock on the elevated \textsc{[Cii]} emission.
Recently some work has suggested that mechanical heating due to turbulence in shocks can contribute to \textsc{[Cii]} emission at high redshift (e.g., \citealt{2010ApJ...724..957S}; \citealt{2013A&A...550A.106L}; \citealt{2013ApJ...777...66A}; \citealt{2015ApJ...799...13B}).
For instance, \citet{2013ApJ...777...66A} reported that the resolved shocked regions of Stephan's Quintet have exceptionally high \textsc{[Cii]}--FIR ratio and they also suggest that this could be commonplace for high-redshift galaxies. 
\citet{2015ApJ...799...13B} suggested that a variety of shocks, originating from major-merger, intergalactic gas accretion, and stellar outflows, might contribute to the elevated \textsc{[Cii]} emission.
ALMA3 shows complicated rest-frame UV morphologies and \textsc{[Cii]} velocity structures (Fig.~\ref{cii_map}), which is suggestive of galaxy-galaxy interaction (dust obscuration may also contribute to it).
ALMA3 hosts intense star-formation activity, as the dust continuum detection shows, 
and appears to be a relatively evolved system with large stellar mass $M_*\approx10^{11} M_\odot$ 
(\citealt{2015ApJ...799...38K}) comparable to the derived dynamical mass (we need to recognize both estimates contain large uncertainties).
Therefore galactic outflow may interact with intergalactic gas stream (e.g., \citealt{2016arXiv160904405C}).
Thus shock heating might be a contributor of \textsc{[Cii]} emission from ALMA3.


One key question is the role of environment, since ALMA3 is located within a giant LAB, SSA22-LAB1. 
LAB1 resides in a remarkable proto-cluster and is associated with a number of star-forming galaxies, which may reflect the abundant gas accretion from cosmic web. The overdensity of galaxies may lead a high frequency of galaxy-galaxy interaction. Therefore the unique environment might account for the relatively strong \textsc{[Cii]} line. On the other hand, if ALMA1 and ALMA2 are actually at redshifts similar to confirmed LAB1 members, the absence of detectable \textsc{[Cii]} would mean diversity of the ISM state within a LAB.
While we detected the \textsc{[Cii]} line from a massive, dusty star-forming galaxy,
much deeper observations of FIR lines like \textsc{[Cii]} and \textsc{[Nii]} toward a giant LAB at $z\sim3$, which allows us to assess the ISM state in UV/optical selected  galaxies (e.g., LBGs like C11 in LAB1), is highly expected.
Such surveys will give us an opportunity to estimate how the ISM in the galaxies evolve in biased regions in the early universe, through the comparison with other FIR line observations of galaxies in a biased region (e.g., AzTEC3 and LBG1 at $z=5.3$; e.g., \citealt{2014ApJ...796...84R}; \citealt{2016arXiv160702520P}) or galaxies in general environment in the same era.
 


\acknowledgments
We greatly appreciate the anonymous referee for a helpful report.
HU is supported by the ALMA Japan Research Grant of NAOJ Chile Observatory, NAOJ-ALMA-0071, 0131, 140, and 0152. HU is supported by JSPS Grant-in-Aid for Research Activity Start-up (16H06713).
HU is thankful for the support from JSPS KAKENHI No 16H02166 (PI. Y. Taniguchi). 
YT is supported by JSPS KAKENHI No. 25102073.
RJI acknowledges support from ERC in the form of the Advanced Investigator Programme, 321302, COSMICISM.
IRS acknowledge support from STFC (ST/L00075X/1).
IRS acknowledge support from the ERC Advanced Investigator program
DUSTYGAL 321334, and a Royal Society/Wolfson Merit Award. 
MH acknowledges the support of the Swedish Research Council,
Vetenskapsr{\aa}det and the Swedish National Space Board (SNSB), and is Fellow of the Knut and Alice Wallenberg Foundation.
This paper makes use of the following ALMA data: ADS/JAO.ALMA\#2013.1.00159.S, ADS/JAO.ALMA\#2013.1.00704.S. ALMA is a partnership of ESO (representing its member states), NSF (USA) and NINS (Japan), together with NRC (Canada) and NSC and ASIAA (Taiwan) and KASI (Republic of Korea), in cooperation with the Republic of Chile. The Joint ALMA Observatory is operated by ESO, AUI/NRAO and NAOJ.



{\it Facilities:} \facility{ALMA}.

\bibliographystyle{apj}

\begin{thebibliography}{}
\expandafter\ifx\csname natexlab\endcsname\relax\def\natexlab#1{#1}\fi

\bibitem[{{Appleton} {et~al.}(2013){Appleton}, {Guillard}, {Boulanger},
  {Cluver}, {Ogle}, {Falgarone}, {Pineau des For{\^e}ts}, {O'Sullivan}, {Duc},
  {Gallagher}, {Gao}, {Jarrett}, {Konstantopoulos}, {Lisenfeld}, {Lord}, {Lu},
  {Peterson}, {Struck}, {Sturm}, {Tuffs}, {Valchanov}, {van der Werf}, \&
  {Xu}}]{2013ApJ...777...66A}
{Appleton}, P.~N., {Guillard}, P., {Boulanger}, F., {et~al.} 2013, \apj, 777,
  66

\bibitem[{{B{\'e}thermin} {et~al.}(2016){B{\'e}thermin}, {De Breuck},
  {Gullberg}, {Aravena}, {Bothwell}, {Chapman}, {Gonzalez}, {Greve}, {Litke},
  {Ma}, {Malkan}, {Marrone}, {Murphy}, {Spilker}, {Stark}, {Strandet},
  {Vieira}, {Wei{\ss}}, \& {Welikala}}]{2016A&A...586L...7B}
{B{\'e}thermin}, M., {De Breuck}, C., {Gullberg}, B., {et~al.} 2016, \aap, 586,
  L7

\bibitem[{{Brisbin} {et~al.}(2015){Brisbin}, {Ferkinhoff}, {Nikola},
  {Parshley}, {Stacey}, {Spoon}, {Hailey-Dunsheath}, \&
  {Verma}}]{2015ApJ...799...13B}
{Brisbin}, D., {Ferkinhoff}, C., {Nikola}, T., {et~al.} 2015, \apj, 799, 13

\bibitem[{{Capak} {et~al.}(2015){Capak}, {Carilli}, {Jones}, {Casey},
  {Riechers}, {Sheth}, {Carollo}, {Ilbert}, {Karim}, {Lefevre}, {Lilly},
  {Scoville}, {Smolcic}, \& {Yan}}]{2015Natur.522..455C}
{Capak}, P.~L., {Carilli}, C., {Jones}, G., {et~al.} 2015, \nat, 522, 455

\bibitem[{{Chapman} {et~al.}(2004){Chapman}, {Scott}, {Windhorst}, {Frayer},
  {Borys}, {Lewis}, \& {Ivison}}]{2004ApJ...606...85C}
{Chapman}, S.~C., {Scott}, D., {Windhorst}, R.~A., {et~al.} 2004, \apj, 606, 85

\bibitem[{{Chapman} {et~al.}(2003){Chapman}, {Windhorst}, {Odewahn}, {Yan}, \&
  {Conselice}}]{2003ApJ...599...92C}
{Chapman}, S.~C., {Windhorst}, R., {Odewahn}, S., {Yan}, H., \& {Conselice}, C.
  2003, \apj, 599, 92

\bibitem[{{Cicone} {et~al.}(2015){Cicone}, {Maiolino}, {Gallerani}, {Neri},
  {Ferrara}, {Sturm}, {Fiore}, {Piconcelli}, \&
  {Feruglio}}]{2015A&A...574A..14C}
{Cicone}, C., {Maiolino}, R., {Gallerani}, S., {et~al.} 2015, \aap, 574, A14

\bibitem[{{Cornuault} {et~al.}(2016){Cornuault}, {Lehnert}, {Boulanger}, \&
  {Guillard}}]{2016arXiv160904405C}
{Cornuault}, N., {Lehnert}, M., {Boulanger}, F., \& {Guillard}, P. 2016, ArXiv
  e-prints, arXiv:1609.04405

\bibitem[{{De Breuck} {et~al.}(2014){De Breuck}, {Williams}, {Swinbank},
  {Caselli}, {Coppin}, {Davis}, {Maiolino}, {Nagao}, {Smail}, {Walter},
  {Wei{\ss}}, \& {Zwaan}}]{2014A&A...565A..59D}
{De Breuck}, C., {Williams}, R.~J., {Swinbank}, M., {et~al.} 2014, \aap, 565,
  A59

\bibitem[{{Decarli} {et~al.}(2014){Decarli}, {Walter}, {Carilli}, {Bertoldi},
  {Cox}, {Ferkinhoff}, {Groves}, {Maiolino}, {Neri}, {Riechers}, \&
  {Weiss}}]{2014ApJ...782L..17D}
{Decarli}, R., {Walter}, F., {Carilli}, C., {et~al.} 2014, \apjl, 782, L17

\bibitem[{{D{\'{\i}}az-Santos} {et~al.}(2013){D{\'{\i}}az-Santos}, {Armus},
  {Charmandaris}, {Stierwalt}, {Murphy}, {Haan}, {Inami}, {Malhotra},
  {Meijerink}, {Stacey}, {Petric}, {Evans}, {Veilleux}, {van der Werf}, {Lord},
  {Lu}, {Howell}, {Appleton}, {Mazzarella}, {Surace}, {Xu}, {Schulz},
  {Sanders}, {Bridge}, {Chan}, {Frayer}, {Iwasawa}, {Melbourne}, \&
  {Sturm}}]{2013ApJ...774...68D}
{D{\'{\i}}az-Santos}, T., {Armus}, L., {Charmandaris}, V., {et~al.} 2013, \apj,
  774, 68

\bibitem[{{Dijkstra} \& {Loeb}(2009)}]{2009MNRAS.400.1109D}
{Dijkstra}, M., \& {Loeb}, A. 2009, \mnras, 400, 1109

\bibitem[{{Engel} {et~al.}(2010){Engel}, {Tacconi}, {Davies}, {Neri}, {Smail},
  {Chapman}, {Genzel}, {Cox}, {Greve}, {Ivison}, {Blain}, {Bertoldi}, \&
  {Omont}}]{2010ApJ...724..233E}
{Engel}, H., {Tacconi}, L.~J., {Davies}, R.~I., {et~al.} 2010, \apj, 724, 233

\bibitem[{{Erb} {et~al.}(2011){Erb}, {Bogosavljevi{\'c}}, \&
  {Steidel}}]{2011ApJ...740L..31E}
{Erb}, D.~K., {Bogosavljevi{\'c}}, M., \& {Steidel}, C.~C. 2011, \apjl, 740,
  L31

\bibitem[{{Geach} {et~al.}(2005){Geach}, {Matsuda}, {Smail}, {Chapman},
  {Yamada}, {Ivison}, {Hayashino}, {Ohta}, {Shioya}, \&
  {Taniguchi}}]{2005MNRAS.363.1398G}
{Geach}, J.~E., {Matsuda}, Y., {Smail}, I., {et~al.} 2005, \mnras, 363, 1398

\bibitem[{{Geach} {et~al.}(2014){Geach}, {Bower}, {Alexander}, {Blain},
  {Bremer}, {Chapin}, {Chapman}, {Clements}, {Coppin}, {Dunlop}, {Farrah},
  {Jenness}, {Koprowski}, {Micha{\l}owski}, {Robson}, {Scott}, {Smith},
  {Spaans}, {Swinbank}, \& {van der Werf}}]{2014ApJ...793...22G}
{Geach}, J.~E., {Bower}, R.~G., {Alexander}, D.~M., {et~al.} 2014, \apj, 793,
  22

\bibitem[{{Geach} {et~al.}(2016){Geach}, {Narayanan}, {Matsuda}, {Hayes},
  {Mas-Ribas}, {Dijkstra}, {Steidel}, {Chapman}, {Feldmann}, {Avison},
  {Agertz}, {Ao}, {Birkinshaw}, {Bremer}, {Clements}, {Dannerbauer}, {Farrah},
  {Harrison}, {Kubo}, {Micha{\l}owski}, {Scott}, {Smith}, {Spaans}, {Simpson},
  {Swinbank}, {Taniguchi}, {van der Werf}, {Verma}, \&
  {Yamada}}]{2016ApJ...832...37G}
{Geach}, J.~E., {Narayanan}, D., {Matsuda}, Y., {et~al.} 2016, \apj, 832, 37

\bibitem[{{Gullberg} {et~al.}(2015){Gullberg}, {De Breuck}, {Vieira},
  {Wei{\ss}}, {Aguirre}, {Aravena}, {B{\'e}thermin}, {Bradford}, {Bothwell},
  {Carlstrom}, {Chapman}, {Fassnacht}, {Gonzalez}, {Greve}, {Hezaveh},
  {Holzapfel}, {Husband}, {Ma}, {Malkan}, {Marrone}, {Menten}, {Murphy},
  {Reichardt}, {Spilker}, {Stark}, {Strandet}, \&
  {Welikala}}]{2015MNRAS.449.2883G}
{Gullberg}, B., {De Breuck}, C., {Vieira}, J.~D., {et~al.} 2015, \mnras, 449,
  2883

\bibitem[{{Hayes} {et~al.}(2011){Hayes}, {Scarlata}, \&
  {Siana}}]{2011Natur.476..304H}
{Hayes}, M., {Scarlata}, C., \& {Siana}, B. 2011, \nat, 476, 304

\bibitem[{{Ikarashi} {et~al.}(2015){Ikarashi}, {Ivison}, {Caputi}, {Aretxaga},
  {Dunlop}, {Hatsukade}, {Hughes}, {Iono}, {Izumi}, {Kawabe}, {Kohno}, {Lagos},
  {Motohara}, {Nakanishi}, {Ohta}, {Tamura}, {Umehata}, {Wilson}, {Yabe}, \&
  {Yun}}]{2015ApJ...810..133I}
{Ikarashi}, S., {Ivison}, R.~J., {Caputi}, K.~I., {et~al.} 2015, \apj, 810, 133

\bibitem[{{Inoue} {et~al.}(2016){Inoue}, {Tamura}, {Matsuo}, {Mawatari},
  {Shimizu}, {Shibuya}, {Ota}, {Yoshida}, {Zackrisson}, {Kashikawa}, {Kohno},
  {Umehata}, {Hatsukade}, {Iye}, {Matsuda}, {Okamoto}, \&
  {Yamaguchi}}]{2016Sci...352.1559I}
{Inoue}, A.~K., {Tamura}, Y., {Matsuo}, H., {et~al.} 2016, Science, 352, 1559

\bibitem[{{Israel} {et~al.}(1996){Israel}, {Bontekoe}, \&
  {Kester}}]{1996A&A...308..723I}
{Israel}, F.~P., {Bontekoe}, T.~R., \& {Kester}, D.~J.~M. 1996, \aap, 308, 723

\bibitem[{{Kubo} {et~al.}(2015){Kubo}, {Yamada}, {Ichikawa}, {Kajisawa},
  {Matsuda}, \& {Tanaka}}]{2015ApJ...799...38K}
{Kubo}, M., {Yamada}, T., {Ichikawa}, T., {et~al.} 2015, \apj, 799, 38

\bibitem[{{Kubo} {et~al.}(2013){Kubo}, {Uchimoto}, {Yamada}, {Kajisawa},
  {Ichikawa}, {Matsuda}, {Akiyama}, {Hayashino}, {Konishi}, {Nishimura},
  {Omata}, {Suzuki}, {Tanaka}, {Yoshikawa}, {Alexander}, {Fazio}, {Huang}, \&
  {Lehmer}}]{2013ApJ...778..170K}
{Kubo}, M., {Uchimoto}, Y.~K., {Yamada}, T., {et~al.} 2013, \apj, 778, 170

\bibitem[{{Lesaffre} {et~al.}(2013){Lesaffre}, {Pineau des For{\^e}ts},
  {Godard}, {Guillard}, {Boulanger}, \& {Falgarone}}]{2013A&A...550A.106L}
{Lesaffre}, P., {Pineau des For{\^e}ts}, G., {Godard}, B., {et~al.} 2013, \aap,
  550, A106

\bibitem[{{Madden} {et~al.}(1993){Madden}, {Geis}, {Genzel}, {Herrmann},
  {Jackson}, {Poglitsch}, {Stacey}, \& {Townes}}]{1993ApJ...407..579M}
{Madden}, S.~C., {Geis}, N., {Genzel}, R., {et~al.} 1993, \apj, 407, 579

\bibitem[{{Matsuda} {et~al.}(2004){Matsuda}, {Yamada}, {Hayashino}, {Tamura},
  {Yamauchi}, {Ajiki}, {Fujita}, {Murayama}, {Nagao}, {Ohta}, {Okamura},
  {Ouchi}, {Shimasaku}, {Shioya}, \& {Taniguchi}}]{2004AJ....128..569M}
{Matsuda}, Y., {Yamada}, T., {Hayashino}, T., {et~al.} 2004, \aj, 128, 569

\bibitem[{{Matsuda} {et~al.}(2005){Matsuda}, {Yamada}, {Hayashino}, {Tamura},
  {Yamauchi}, {Murayama}, {Nagao}, {Ohta}, {Okamura}, {Ouchi}, {Shimasaku},
  {Shioya}, \& {Taniguchi}}]{2005ApJ...634L.125M}
---. 2005, \apjl, 634, L125

\bibitem[{{McLinden} {et~al.}(2013){McLinden}, {Malhotra}, {Rhoads}, {Hibon},
  {Weijmans}, \& {Tilvi}}]{2013ApJ...767...48M}
{McLinden}, E.~M., {Malhotra}, S., {Rhoads}, J.~E., {et~al.} 2013, \apj, 767,
  48

\bibitem[{{McMullin} {et~al.}(2007){McMullin}, {Waters}, {Schiebel}, {Young},
  \& {Golap}}]{2007ASPC..376..127M}
{McMullin}, J.~P., {Waters}, B., {Schiebel}, D., {Young}, W., \& {Golap}, K.
  2007, in Astronomical Society of the Pacific Conference Series, Vol. 376,
  Astronomical Data Analysis Software and Systems XVI, ed. R.~A. {Shaw},
  F.~{Hill}, \& D.~J. {Bell}, 127

\bibitem[{{Mori} \& {Umemura}(2006)}]{2006Natur.440..644M}
{Mori}, M., \& {Umemura}, M. 2006, \nat, 440, 644

\bibitem[{{Nagao} {et~al.}(2012){Nagao}, {Maiolino}, {De Breuck}, {Caselli},
  {Hatsukade}, \& {Saigo}}]{2012A&A...542L..34N}
{Nagao}, T., {Maiolino}, R., {De Breuck}, C., {et~al.} 2012, \aap, 542, L34

\bibitem[{{Nagao} {et~al.}(2011){Nagao}, {Maiolino}, {Marconi}, \&
  {Matsuhara}}]{2011A&A...526A.149N}
{Nagao}, T., {Maiolino}, R., {Marconi}, A., \& {Matsuhara}, H. 2011, \aap, 526,
  A149

\bibitem[{{Oberst} {et~al.}(2006){Oberst}, {Parshley}, {Stacey}, {Nikola},
  {L{\"o}hr}, {Harnett}, {Tothill}, {Lane}, {Stark}, \&
  {Tucker}}]{2006ApJ...652L.125O}
{Oberst}, T.~E., {Parshley}, S.~C., {Stacey}, G.~J., {et~al.} 2006, \apjl, 652,
  L125

\bibitem[{{Pavesi} {et~al.}(2016){Pavesi}, {Riechers}, {Capak}, {Carilli},
  {Sharon}, {Stacey}, {Karim}, {Scoville}, \& {Smolcic}}]{2016arXiv160702520P}
{Pavesi}, R., {Riechers}, D.~A., {Capak}, P.~L., {et~al.} 2016, ArXiv e-prints,
  arXiv:1607.02520

\bibitem[{{Rawle} {et~al.}(2014){Rawle}, {Egami}, {Bussmann}, {Gurwell},
  {Ivison}, {Boone}, {Combes}, {Danielson}, {Rex}, {Richard}, {Smail},
  {Swinbank}, {Altieri}, {Blain}, {Clement}, {Dessauges-Zavadsky}, {Edge},
  {Fazio}, {Jones}, {Kneib}, {Omont}, {P{\'e}rez-Gonz{\'a}lez}, {Schaerer},
  {Valtchanov}, {van der Werf}, {Walth}, {Zamojski}, \&
  {Zemcov}}]{2014ApJ...783...59R}
{Rawle}, T.~D., {Egami}, E., {Bussmann}, R.~S., {et~al.} 2014, \apj, 783, 59

\bibitem[{{Riechers} {et~al.}(2014){Riechers}, {Carilli}, {Capak}, {Scoville},
  {Smol{\v c}i{\'c}}, {Schinnerer}, {Yun}, {Cox}, {Bertoldi}, {Karim}, \&
  {Yan}}]{2014ApJ...796...84R}
{Riechers}, D.~A., {Carilli}, C.~L., {Capak}, P.~L., {et~al.} 2014, \apj, 796,
  84

\bibitem[{{Simpson} {et~al.}(2015){Simpson}, {Smail}, {Swinbank}, {Almaini},
  {Blain}, {Bremer}, {Chapman}, {Chen}, {Conselice}, {Coppin}, {Danielson},
  {Dunlop}, {Edge}, {Farrah}, {Geach}, {Hartley}, {Ivison}, {Karim}, {Lani},
  {Ma}, {Meijerink}, {Micha{\l}owski}, {Mortlock}, {Scott}, {Simpson},
  {Spaans}, {Thomson}, {van Kampen}, \& {van der Werf}}]{2015ApJ...799...81S}
{Simpson}, J.~M., {Smail}, I., {Swinbank}, A.~M., {et~al.} 2015, \apj, 799, 81

\bibitem[{{Stacey} {et~al.}(1991){Stacey}, {Geis}, {Genzel}, {Lugten},
  {Poglitsch}, {Sternberg}, \& {Townes}}]{1991ApJ...373..423S}
{Stacey}, G.~J., {Geis}, N., {Genzel}, R., {et~al.} 1991, \apj, 373, 423

\bibitem[{{Stacey} {et~al.}(2010){Stacey}, {Hailey-Dunsheath}, {Ferkinhoff},
  {Nikola}, {Parshley}, {Benford}, {Staguhn}, \&
  {Fiolet}}]{2010ApJ...724..957S}
{Stacey}, G.~J., {Hailey-Dunsheath}, S., {Ferkinhoff}, C., {et~al.} 2010, \apj,
  724, 957

\bibitem[{{Steidel} {et~al.}(2000){Steidel}, {Adelberger}, {Shapley},
  {Pettini}, {Dickinson}, \& {Giavalisco}}]{2000ApJ...532..170S}
{Steidel}, C.~C., {Adelberger}, K.~L., {Shapley}, A.~E., {et~al.} 2000, \apj,
  532, 170

\bibitem[{{Steidel} {et~al.}(2003){Steidel}, {Adelberger}, {Shapley},
  {Pettini}, {Dickinson}, \& {Giavalisco}}]{2003ApJ...592..728S}
---. 2003, \apj, 592, 728

\bibitem[{{Swinbank} {et~al.}(2014){Swinbank}, {Simpson}, {Smail}, {Harrison},
  {Hodge}, {Karim}, {Walter}, {Alexander}, {Brandt}, {de Breuck}, {da Cunha},
  {Chapman}, {Coppin}, {Danielson}, {Dannerbauer}, {Decarli}, {Greve},
  {Ivison}, {Knudsen}, {Lagos}, {Schinnerer}, {Thomson}, {Wardlow}, {Wei{\ss}},
  \& {van der Werf}}]{2014MNRAS.438.1267S}
{Swinbank}, A.~M., {Simpson}, J.~M., {Smail}, I., {et~al.} 2014, \mnras, 438,
  1267

\bibitem[{{Taniguchi} \& {Shioya}(2000)}]{2000ApJ...532L..13T}
{Taniguchi}, Y., \& {Shioya}, Y. 2000, \apjl, 532, L13

\bibitem[{{Uchimoto} {et~al.}(2012){Uchimoto}, {Yamada}, {Kajisawa}, {Kubo},
  {Ichikawa}, {Matsuda}, {Akiyama}, {Hayashino}, {Konishi}, {Nishimura},
  {Omata}, {Suzuki}, {Tanaka}, {Tokoku}, \& {Yoshikawa}}]{2012ApJ...750..116U}
{Uchimoto}, Y.~K., {Yamada}, T., {Kajisawa}, M., {et~al.} 2012, \apj, 750, 116

\bibitem[{{Umehata} {et~al.}(2015){Umehata}, {Tamura}, {Kohno}, {Ivison},
  {Alexander}, {Geach}, {Hatsukade}, {Hughes}, {Ikarashi}, {Kato}, {Izumi},
  {Kawabe}, {Kubo}, {Lee}, {Lehmer}, {Makiya}, {Matsuda}, {Nakanishi}, {Saito},
  {Smail}, {Yamada}, {Yamaguchi}, \& {Yun}}]{2015ApJ...815L...8U}
{Umehata}, H., {Tamura}, Y., {Kohno}, K., {et~al.} 2015, \apjl, 815, L8

\bibitem[{{Umehata} {et~al.}(2016){Umehata}, {Tamura}, {Kohno}, {Ivison},
  {Smail}, {Hatsukade}, {Nakanishi}, {Kato}, {Ikarashi}, {Matsuda}, {Fujimoto},
  {Iono}, {Lee}, {Steidel}, {Saito}, {Alexander}, {Yun}, \&
  {Kubo}}]{2016arXiv161109857U}
---. 2016, ArXiv e-prints, arXiv:1611.09857

\bibitem[{{Yang} {et~al.}(2009){Yang}, {Zabludoff}, {Tremonti}, {Eisenstein},
  \& {Dav{\'e}}}]{2009ApJ...693.1579Y}
{Yang}, Y., {Zabludoff}, A., {Tremonti}, C., {Eisenstein}, D., \& {Dav{\'e}},
  R. 2009, \apj, 693, 1579

\bibitem[{{Zhao} {et~al.}(2016){Zhao}, {Lu}, {Xu}, {Gao}, {Lord},
  {Charmandaris}, {Diaz-Santos}, {Evans}, {Howell}, {Petric}, {van der Werf},
  \& {Sanders}}]{2016ApJ...819...69Z}
{Zhao}, Y., {Lu}, N., {Xu}, C.~K., {et~al.} 2016, \apj, 819, 69

\end{thebibliography}





\end{document}